\documentclass[useAMS,usenatbib,letterpaper]{mn2e}
\usepackage[varg]{txfonts}
\usepackage{graphicx}
\usepackage{natbib}
\usepackage{hyperref}
\hypersetup{
  colorlinks   = true,    
  urlcolor     = blue,    
  linkcolor    = blue,    
  citecolor    = blue      
}
\usepackage[usenames,dvipsnames]{xcolor}

\def\apjl{ApJL }

\def\aj{AJ }

\def\apj{ApJ }
\def\pasp{PASP }

\def\apjs{ApJS }

\def\mnras{MNRAS }
\def\aap{A\&A }
\def\aaps{A\&A Suppl }

\newcommand{\pmra}{$\mu_{\alpha *}$}
\newcommand{\pmdec}{$\mu_{\delta}$}

\begin{document}

\title[Discovery of a low-mass companion to HD 984]
{Discovery of a Low-Mass Companion to the F7V star HD 984}

\author[Meshkat et al.]
  {T.~Meshkat,$^1$\thanks{Based on observations collected at the European
  Organization for Astronomical Research in the Southern Hemisphere,
  Chile, ESO under program numbers 089.C-0617(A), 089.C-0149(A), 093.C-0626(A)}
  M. Bonnefoy,$^{2,3}$ E.~E. Mamajek,$^4$ S. P. Quanz,$^5$, G. Chauvin,$^{2,3}$ \newauthor
  M. A. Kenworthy,$^1$ J. Rameau,$^6$ M. R. Meyer,$^5$ A.-M. Lagrange,$^{2,3}$
  J. Lannier,$^{2,3}$ \newauthor P. Delorme$^{2,3}$\\
  $^1$Leiden Observatory, P.O. Box 9513, Niels Bohrweg 2,
  2300 RA Leiden, The Netherlands\\
  $^2$Universit\'{e} Grenoble Alpes, IPAG, 38000 Grenoble, France\\
  $^3$CNRS, IPAG, 38000 Grenoble, France\\
  $^4$Department of Physics and Astronomy, University of
  Rochester, Rochester, NY 14627-0171, USA\\
  $^5$Institute for Astronomy, ETH Zurich,
  Wolfgang-Pauli-Strasse 27, 8093 Zurich, Switzerland\\
  $^6$Institut pour la Recherche sur les Exoplan\`etes (IREX), D\'epartement de physique, Universit\'e de Montr\'eal, C.P. 6128 
 Succ. Centre-ville, Montr\'eal, QC H3C 3J7, Canada}
\date{2015 March 4}

\pagerange{\pageref{firstpage}--\pageref{lastpage}} \pubyear{2015}

\maketitle

\label{firstpage}

\begin{abstract}
We report the discovery of a low-mass companion to the nearby ($d$ = 47 pc) F7V star
HD 984. The companion is detected $0\farcs19$ away from its host star in the $L'$ band 
with the Apodizing Phase Plate on NaCo/VLT and was recovered by $L'$-band 
non-coronagraphic imaging data taken a few days later. We confirm the companion is 
co-moving with the star with SINFONI integral field spectrograph $H+K$ data. We 
present the first published data obtained with SINFONI in pupil-tracking mode. HD 984 
has been argued to be a kinematic member of the 30 Myr-old Columba group, and its 
HR diagram position is not altogether inconsistent with being a ZAMS star of this age. 
By consolidating different age indicators, including isochronal age, coronal X-ray 
emission, and stellar rotation, we independently estimate a main sequence age of 
115\,$\pm$\,85 Myr (95\% CL) which does not rely on this kinematic association. The 
mass of directly imaged companions are usually inferred from theoretical evolutionary 
tracks, which are highly dependent on the age of the star. Based on the age extrema, 
we demonstrate that with our photometric data alone, the companion's mass is highly 
uncertain: between 33 and 96 M$_{\rm Jup}$ (0.03-0.09 M$_{\odot}$) using the COND 
evolutionary models. We compare the companion's SINFONI spectrum with field dwarf 
spectra to break this degeneracy. Based on the slope and shape of the spectrum in the 
$H$-band, we conclude that the companion is an M$6.0\pm0.5$ dwarf. The age of the 
system is not further constrained by the companion, as M dwarfs are poorly fit on 
low-mass evolutionary tracks. This discovery emphasizes the importance of obtaining a 
spectrum to spectral type companions around F-stars.
\end{abstract}


\section{Introduction}

Young stars are the primary targets of exoplanet imaging surveys because associated 
planets are warm and therefore bright in the infrared. A handful of brown dwarfs and 
low-mass stellar companions have been found in these surveys (PZ Tel B; 
\citealt{Biller10}, CD-35 2722 B; \citealt{Wahhaj11}, HD 1160 B,C; \citealt{Nielsen12}). 
Masses of directly imaged companions are estimated from the companion's luminosity 
and theoretical evolutionary models, which are very sensitive to the age and distance 
of the host star.

The companion mass-ratio distribution (CMRD) quantifies the mass ratio of a binary 
system \citep{Reggiani13}. Based on observational data, the initial mass function (IMF) 
for brown dwarfs and very-low-mass-stars (0.08-0.2$M_{\odot}$) likely differs from 
stars \citep{Thies15}. The primary formation mechanisms for these low-mass 
companions, including fragmentation and capture, is still under debate, making each 
new discovered low-mass companion an important test case for the theoretical 
formation mechanisms of low-mass companion. The orbital motion of the companion 
around the primary star, measured as a small arc on the sky, can be used to find 
orbital solutions \citep{Pearce15}. Some orbital properties, such as eccentricity, can 
be constrained with only two measurements \citep{Biller10}. Additionally, with multiple 
epoch astrometric measurements from direct imaging and radial velocity data, the 
dynamical mass of a companion can be measured. This acts as an important 
comparison with the inferred companion masses from theoretical evolutionary models 
\citep{Bonnefoy09,Dupuy15,Close07}.

We report the detection of a companion around the F7V star HD 984 (HIP 1134). This 
star has been part of many imaging surveys searching for planets 
\citep{Brandt14,Rameau13} due to its proximity, brightness 
\citep[$d$ = 47.1\,$\pm$\,1.4 pc; V\,=\,7.3;][]{ESA97,vanLeeuwen07}, and proposed 
youth (30 Myr, \citealt{Zuckerman11}). However, no comoving companions have yet 
been reported. Ground-based sub-mm and {\it Spitzer} infrared photometry of HD 984 
have not detected any evidence of a dusty debris disk 
\citep{Mamajek04, Carpenter09, Ballering13}.

In Section \ref{sec:obs} we describe our coronagraphic and non-coronagraphic 
observations with NaCo on the VLT, and our SINFONI/VLT integral field spectrographic 
data. In Section \ref{sec:phot} we measure the companion's photometry and 
astrometry. In Section \ref{sec:Analysis} we discuss previous age estimates of HD 984, 
estimate the age of the primary based on its main sequence isochronal age and other 
age indicators, and derive the mass of the companion based on evolutionary models 
and spectral analysis. We conclude in Section \ref{sec:Conclusion}.

\section{Observations}
\label{sec:obs}
\subsection{NaCo/VLT}

Observations of HD 984 were taken on UT 2012 July 18 and 20 (089.C-0617(A), 
PI: Sascha Quanz) at the Very Large Telescope (VLT)/UT4 with NaCo 
(\citealt{Lenzen03,Rousset03}). The Apodizing Phase Plate coronagraph 
(APP; \citealt{Kenworthy10,Quanz10}) was used for  diffraction suppression thus 
increasing the chances of detecting a companion very close to the target star. Data 
were obtained with the L27 camera, in the $L'$-band filter ($\lambda$ = 3.80$\mu$m 
and $\Delta\lambda$\,=\,0.62$\mu$m). The visible wavefront sensor was used with 
HD 984 as the natural guide star. We observed in pupil tracking mode 
\citep{Kasper09} for Angular Differential Imaging (ADI; \citealt{Marois06}). We 
intentionally saturated the PSF core to increase the signal-to-noise from potential
companions in each exposure. Unsaturated data were also obtained to calibrate 
photometry relative to the central star.

The APP suppresses diffraction over a 180$^{\circ}$ wedge on one side of the target 
star. Excess scattered light is increased on the other side of the target that is not 
used in the data analysis. Two datasets were obtained with different initial position 
angles (P.A.) for full 360$^{\circ}$ coverage around the target star. The field rotation 
was 47$^{\circ}$.4 in the first hemisphere and 42$^{\circ}$.5 in the second 
hemisphere.

Direct imaging observations of HD 984 were obtained on VLT/NaCo on UT 2012 July 
20 (089.C-0149(A), PI: Julien Rameau).  The data were taken with the L27 camera on 
NaCo in ADI pupil tracking mode. Both the saturated and unsaturated data were 
imaged in $L'$-band with the same exposure time, but in the unsaturated images a 
neutral density filter (ND\_LONG) was used. The field rotation was 41$^{\circ}$.3 in 
the direct imaging data. All datasets were obtained in cube mode. Each APP cube 
contains 120 frames, with an integration time of 0.5 s per frame. The total integration 
time in the APP was 60 min in hemisphere 1 (60 cubes) and 66 min in hemisphere 2 
(66 cubes). Unsaturated APP exposures were 0.056 s per frame (222 frames in 12 
cubes), with a total time on target of 150 seconds. The direct imaging cubes contain 
100 frames, with 0.2 s per frame, with a total integration time of 48 min for the 
saturated data and 202 seconds for the unsaturated data.

A dither pattern on the detector was used to subtract sky background and detector 
systematics from both datasets, as detailed in \citet{Kenworthy13}. Data cubes are 
subtracted from each other, centroided and averaged over. Optimized principal 
component analysis (PCA) is run on both of the APP hemispheres and direct 
imaging data independently, following \citet{Meshkat14}. Six principal components
are used to model the stellar PSF, which results in the highest signal-to-noise 
detection of the companion. PCA processed frames are derotated and averaged to 
generate the final image with North facing up.

\subsection{SINFONI/VLT}
\label{sec:SINFONI_obs}
Data were obtained on HD 984 with the AO-fed integral field spectrograph SINFONI 
\citep{Eisenhauer03,Bonnet04} at the VLT on UT 2014 September 9 (093.C-0626, 
PI: G. Chauvin) in $\sim0\farcs5$ seeing. The $H+K$ grating was used, which has 
a resolution of $\sim$1500. The spatial sampling was 12.5 mas$\times$25 mas (in 
the horizontal and vertical directions, respectively) resulting in a field of view of 
$0\farcs8\times0\farcs8$. We obtained 42 raw cubes of the target, each consisting 
of 10$\times$4s coadded exposures. Similar to the NaCo observations, data were 
obtained in pupil tracking (PT) mode (Hau et al. in prep). The companion rotated 
49.26$^{\circ}$ around the center of the field of view during our 28 min integration. 

The instrument pipeline version 2.5.2 
\footnote{http://www.eso.org/sci/software/pipelines/sinfoni/sinfoni-pipe-recipes.html} 
was used to correct the raw science frames from hot and non-linear pixels, detector 
gain, and distortion. Final cubes were reconstructed from the resulting frames, 
associated wavelength map, and slitlet positions. Data cubes were corrected for  
OH lines and background emission using a dedicated algorithm \citep{Davies07} 
implemented in the pipeline. 

The wavelength-dependent drift in the star position, caused by the atmospheric 
refraction, was registered (modeled by a 3rd order polynomial) and corrected. We 
took the mean of the parallactic angle values at the beginning and end of a given 
exposure, which is stored in the cube file headers. We processed the cubes with 
the classical-ADI (CADI) algorithm \citep{Marois06} to suppress the stellar flux 
independently at each wavelength (2172 independent spectral resolution elements). 

Data were corrected for telluric absorption lines using the observations of an A3V 
standard star (HD 2811) obtained close in time to HD 984 and at a comparable 
airmass. The resulting spectrum of HD 984 was flux-calibrated using the 2MASS 
$H$-band magnitude of the star \citep{Cutri03} and a spectrum of Vega.

\section{Photometry and Astrometry of HD 984 B} 
\label{sec:phot}

\begin{figure*}
\centering
\includegraphics[width=15cm]{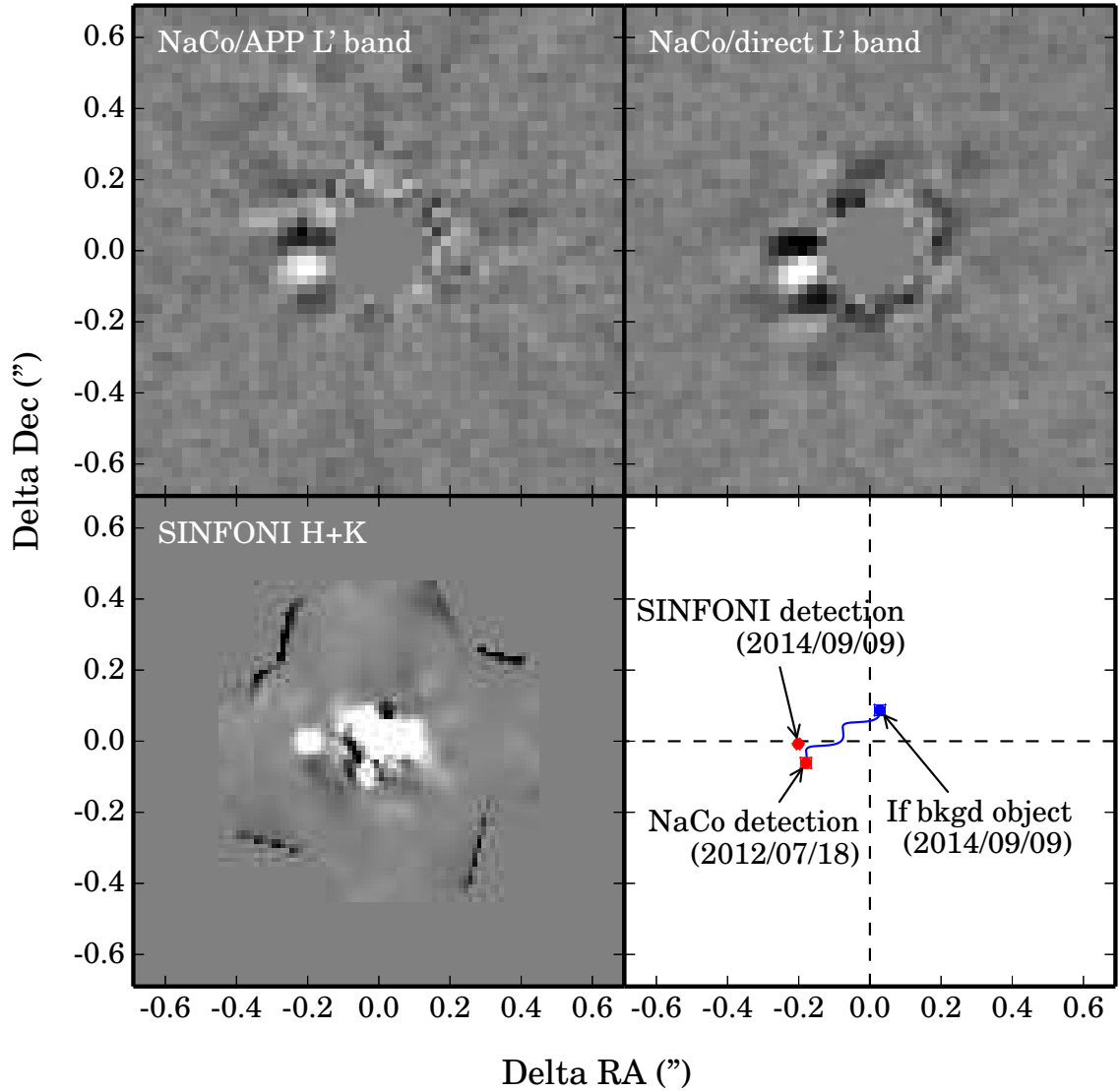}
\caption{Top-left: Final PCA processed image of HD 984 APP hemisphere 1 data with 
North facing up. Twenty principal components were used to model the stellar PSF in 
this image.  
Top-right: Final PCA processed image of HD 984 from direct imaging data with North 
facing up. Six principal components were used to model the stellar PSF. 
Bottom-left: Collapsed $H+K$ SINFONI IFS data cubes processed with CADI, with 
North facing up. All three images are displayed in the same color scale.
Bottom-right: The position of the companion is plotted as red points for both the 
VLT/NaCo dataset epoch (UT 2012 July 18) and the VLT/SINFONI epoch (UT 2014 
September 9). The blue point is the position of the companion if it were a background 
source at the time of the VLT/SINFONI dataset epoch (UT 2014 September 9). Error 
bars are included for all points.}
\label{fig:1}
\end{figure*}

\subsection{NaCo/VLT}
The companion was clearly detected very close to the star (\autoref{fig:1}, APP 
hemisphere 1 on top-left and direct imaging on top-right). The companion detection 
was confirmed using the {\it Pynpoint} pipeline \citep{Amara12} and the IPAG-ADI 
pipeline \citep{Chauvin12}. The detection was robust against changing the number 
of principal components used in the stellar PSF model. 

We used artificial negative companions to determine the astrometry and photometry 
of the companion (following \citealt{Meshkat15}). The unsaturated stellar PSF was
used to generate artificial companions in the APP and direct imaging data. A scaling
factor of 0.018 was applied to the direct imaging unsaturated data to account for the 
attenuation from the neutral density filter.  We injected artificial negative companions 
into the data near the expected position of the companion in steps of 0.1 pixels. The 
artificial companion contrast was varied from 5.0 to 7.0 in steps of 0.01 mag. The 
chi-squared minimization over the $\lambda$/D patch at the location of the artificial 
negative companion yielded the following results.

Based on this analysis, the companion contrast in the APP data is best 
approximated as $\Delta L'$\,=\,6.0\,$\pm$\,0.2 mag 
($L'$\,=\,12.0\,$\pm$\,0.2mag). The angular separation of the companion is
$0\farcs19\,\pm\,0.02$, which corresponds to a projected separation of
9.0$\pm$1.0 AU for the stellar distance of 47.1\,$\pm$\,1.4 pc from
\citet{vanLeeuwen07}. The position angle (P.A.) of the companion is 
$108^{\circ}.8\,\pm\,3^{\circ}.0$. The error in the measurements is
due to the range in artificial companions which successfully subtract the companion
signal.

The companion contrast in the direct imaging data is $\Delta
L'$\,=\,5.9\,$\pm$\,0.3 mag at $0\farcs208\,\pm\,0.023$
(9.8\,$\pm$\,1.1 AU). The P.A. of the companion is $108^{\circ}.9\,\pm\,3^{\circ}.1$ 
corrected to true North orientation, calibrated using $\theta$ Ori, observed at the 
same epoch with the same mode. The $\theta$ Ori stars TCC058, 057, 054, 034, 
and 026 were used to determine the true North 0.41$\pm$0.07$^{\circ}$ with a 
plate scale of 27.11$\pm$0.02 mas \citep{Rameau13}. The astrometry and 
photometry of the companion is in good agreement between the two datasets.  
Since the two datasets were only taken two days apart, we do not consider them 
separate epochs, but rather a confirmation that this companion is not an artifact.

\subsection{SINFONI} 
\begin{figure}
\centering
\includegraphics[width=84mm]{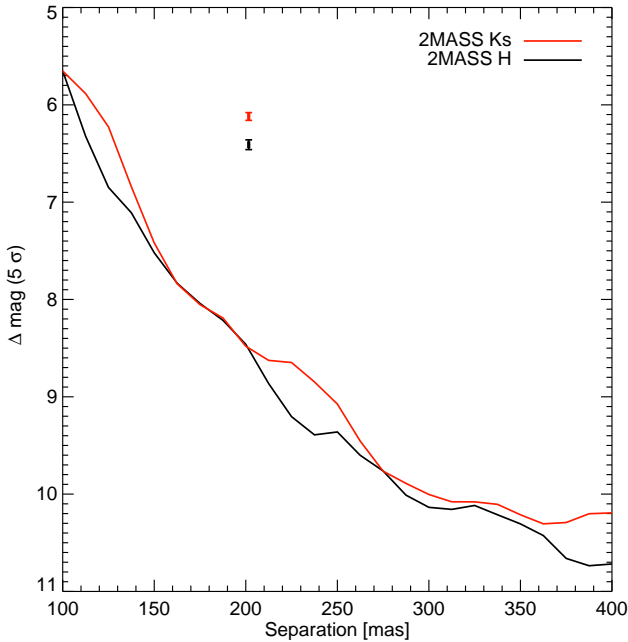}
\caption{Contrast curves for the SINFONI HD 984 $H$ (black curve) and $Ks$ (red 
curve) data, for CADI processing. The companion detection is shown as a point 
with error bars in $H$ (black) and $Ks$ (red), respectively.}
\label{contrast_curve}
\end{figure}

The companion was detected in $H+K$ band IFS SINFONI data (\autoref{fig:1} 
bottom-left). These data are the first published results demonstrating the 
capabilities of the SINFONI instrument in PT mode. The parallactic angle 
associated with each of the 43 data cubes was estimated by taking the mean of 
the parallactic angle at the beginning and the end of an exposure. We applied 
an additional clockwise rotation of 210.92$^{\circ}$ to the frames at the ADI 
reduction step to properly realign the field of view to the North. True North was 
estimated using GQ Lup SINFONI observations on UT 2013 August 24, which 
were calibrated with NaCo observations of the same source on UT 2012 March 
3, assuming orbital motion is negligible between the epochs (following the 
formulae described in the Appendix). 

The companion is detected in all of our ADI analyses. We discovered that 
subtracting the stellar halo of the data cubes collapsed in wavelength followed 
by a realignment of the frame to the North also allowed detection of the 
companion. Unlike ADI processing, this allows for a measurement of the position 
of the companion without the biases associated with the self-subtraction of the 
companion PSF \citep{Bonnefoy11}. We selected 10 frames corresponding to 
the first 10 cubes of the PT sequence and median-combined them after 
realigning to true North. This provided a good removal of residual speckles from 
the stellar halo. We found that  HD 984 B lies at a PA=$92.2\pm0.5^{\circ}$ and 
a separation $\rho=201.6\pm0.4$ mas. The error considers uncertainties in our 
fitting function as well as true North. We also make the assumption that the 
instrument absolute orientation on sky did not vary between our UT 2014 
September 9 observations and the True North calibration (UT 2013 August 24) 
with GQ Lup. The separation assumes a plate scale of 12.5 mas/pixel reported 
in the instrument user manual. The plate scale at the time of HD 984 
observations could not be measured.

HD 984 is a relatively high proper motion star, with \pmra, \pmdec\, =
102.79\,$\pm$\,0.78, -66.36\,$\pm$\,0.36 mas yr$^{-1}$ \citep{vanLeeuwen07}. 
If the companion were a background source, it should be due North of HD 984 
and at a projected separation of $<0\farcs1$ in the SINFONI data (\autoref{fig:1}, 
bottom-right). We estimate that the companion is very unlikely to be a stationary 
background source, with an estimated $\chi^{2}$ probability of less than $10^{-6}$ 
based on the two epoch companion detections, stellar proper motion, distance 
estimates, and the astrometric error on the SINFONI dataset (which is dependent 
on the plate scale). The new P.A. of the companion is consistent with Keplerian 
orbital motion. Based on the astrometry of the companion, we confirm it is bound 
to the star and not a background object. 

The companion flux was integrated in a 6 pixel wide circular aperture in the 
processed CADI cubes to generate the spectrum. We applied the same procedure 
to HD 984. We corrected for flux losses in the companion's spectrum (caused by 
the image processing algorithms) by adding artificial sources. We used the 
unsaturated primary star itself to scale and inject artificial sources in the data. 
Artificial sources were added in each wavelength at the same projected separation 
as the companion, but with a P.A. difference of -90, +90, and +180$^{\circ}$. The 
resulting spectrum was corrected for the telluric lines using the primary spectrum. 
We obtained the final flux-calibrated spectrum of the companion by multiplying the 
flux ratio between the system components and the flux-calibrated spectrum of the 
star. Based on the flux ratio between HD 984 and B in $H$ and $Ks$-band, we 
estimate $H_{2MASS}=12.58 \pm 0.05$ mag and $Ks_{2MASS}=12.19 \pm 0.04$ 
mag for the companion. The spectrum of the companion is analyzed in Section 
\ref{sec:Mass}. 

\autoref{contrast_curve} shows contrast curves comparing the sensitivity to point 
sources in the data processed with the CADI algorithm in $H$ and $Ks$-band. 
The contrast curve was generated following \citet{Chauvin15}. We injected fake 
planets every 10 pixels radially at P.A.s of 0, 120, and 240$^{\circ}$ between 125 
and 375 mas, in order to correct for flux losses. The fake planets are created by 
scaling the flux of the primary star. We created a pixel-to-pixel noise map by 
sliding a box of $5\times5$ pixels from the star to the limit of the SINFONI field of 
view. The $5\sigma$ detection limit is found by dividing the pixel-to-pixel noise 
map by the flux loss, taking into account the relative calibration between the fake 
planet and the primary star.

\section{Analysis}
\label{sec:Analysis}
\subsection{Primary Star HD 984}
 \label{sec:HD984}

HD 984 is a V = 7.32 mag \citep{ESA97} F7V star \citep{Houk99} at a distance of 
47.1\,$\pm$\,1.4 pc \citep[][$\varpi$ = 21.21\,$\pm$\,0.64 mas]{vanLeeuwen07}.
The star has color B-V = 0.522\,$\pm$\,0.010 from the Tycho catalog \citep{ESA97}, 
and for its calculated absolute magnitude (M$_V$ =3.95\,$\pm$\,0.07) the star lies 
squarely on the field star main sequence of \citet{Wright04}, which predicts 
M$_V$(MS) $\simeq$ 3.95 for this B-V color. It is not listed in latest 
version\footnote{8 June 2015 version at http://vizier.cfa.harvard.edu/viz-bin/Cat?B/wds.} 
of the Washington Double Star catalog \citep{Mason01}. The interstellar reddening 
and extinction towards HD\,984 is likely to be negligible as it lies within the Local 
Bubble \citep[e.g.][]{Reis11}. Examining the reddening estimates for 19 stars in 
the catalog of \citet{Reis11} that have Galactic longitude within $\pm$20$^{\circ}$
of HD\,984 and latitude $b$ $<$ -48$^{\circ}$ (their survey only extends to 
$b$ = -60$^{\circ}$), we find no stars within 60\,pc that have a E(b-y) reddening 
that exceeds 1$\sigma$ from zero. Indeed, a fit of distance versus E(b-y) reddening 
for the 19 stars in this region of sky with $d$ $<$ 150 pc is consistent with negligible
reddening (E(b-y) $\simeq$ $d_{pc}$ $\times$ 0.01 mmag). Given this trend, for a 
star at $d$ $\simeq$ 47\,pc, one would expect reddening of E(b-y) $\simeq$ 2 
mmag and extinction $A_V$ $\simeq$ 0.01, i.e. utterly negligible compared to our 
photometric uncertainties. Based on this, we assume negligible reddening or 
extinction for HD\,984 and its companion.

Numerous estimates of the effective temperature have been reported 
\citep[e.g.][]{Valenti05, Masana06, Schroeder09, Casagrande11}.  We adopt the 
recent value from \citet{Casagrande11} (6315\,$\pm$\,89 K), which is also close 
to the median value among recent published estimates. From fitting the 
optical-infrared photometry to the dwarf color sequences from \citet{Pecaut13}, 
we derive an apparent bolometric magnitude of HD 984 to be 
$m_{bol}$ = 7.257\,$\pm$\,0.01 and bolometric flux\footnote{Our F$_{bol}$ 
value is within 3.5\%\, of that of \citet{Casagrande11}: 30.824 pW\,m$^{-2}$.} of
31.94\,$\pm$\,0.29 pW\,m$^{-2}$. Employing the \citet{vanLeeuwen07} parallax, 
this bolometric flux translates to a luminosity of log(L/L$_{\odot}$) = 
0.346\,$\pm$\,0.027 dex. Adopting the \citet{Casagrande11} T$_{\rm eff}$, this 
is consistent with a stellar radius of 1.247\,$\pm$\,0.053 R$_{\odot}$.

\subsubsection{Previous Age Estimates \label{sec:HD984age}}

\citet{Zuckerman11}, \citet{Malo13}, and \citet{Brandt14} consider HD 984 to be 
a member of the Columba group with an age of 30 Myr \citep{Torres08}. 
\citet{Malo13} has demonstrated that HD 984 appears to be comoving with the 
Columba group. However, there remains the possibility that HD 984 could be a
kinematic interloper or that the Columba group is not sufficiently characterized 
to reliably assign ages based on kinematic membership. Isochronal ages of 
$<$0.48 Gyr \citep[68\%CL;][]{Takeda07}, 1.2$^{+0.7}_{-0.9}$ Gyr 
\citep[][]{Valenti05}, and 3.1$^{+1.0}_{-1.6}$ Gyr \citep{Holmberg09} have been 
estimated.

\subsubsection{Isochronal Age \label{sec:isochronalage}}

\begin{figure}
\centering
\includegraphics[width=84mm]{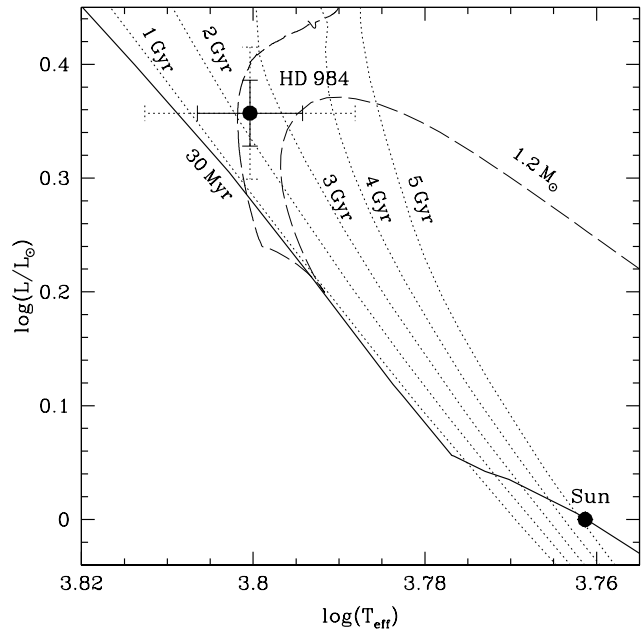}
\caption{HR diagram position of HD 984 and the Sun, with isochrones
  and evolutionary tracks from \citet{Bressan12} overlaid (z = 0.017),
  with 1$\sigma$ solid line and 2$\sigma$ dotted line error bars.  The
  solid line is a 30 Myr isochrone corresponding to the purported age
  of the Columba group. {\it Dotted lines} are isochrones with ages of
  1, 2, 3, 4, 5 Gyr. The {\it dashed line} is a 1.20\,$\pm$\,0.06
  (95\%CL) M$_{\odot}$ track.}
\label{hr}
\end{figure}

The HR diagram position of HD 984 is plotted in \autoref{hr} along with 
isochrones from \citet{Bressan12} (for Z = 0.017). HD 984 is a main sequence 
star, and any derived isochronal age will have large uncertainties. Simulating 
the HR diagram position and interpolating their ages and masses, the 
isochronal ages are consistent with 2.0$^{+2.1}_{-1.8}$ (95\%CL) Gyr and 
1.20\,$\pm$\,0.06 (95\%CL) M$_{\odot}$ (dashed line). However, as 
\autoref{hr} shows, a 2$\sigma$ deviation (dashed error bar line) in both 
T$_{\rm eff}$ and log(L/L$_{\odot}$) are consistent with the 30 Myr isochrone.  
It takes a 1.2\,$M_{\odot}$ star 27 Myr to reach the main sequence, and ages 
of $\lesssim$30 Myr can be ruled out based on the isochronal age constraints 
for the secondary HD\,984\,B (see Section \ref{sec:Mass}). Based on the HR
diagram position of HD 984 compared to stellar evolutionary tracks and 
isochrones, we deduce that the isochronal age of HD 984 is likely between 
$\sim$30 Myr (ZAMS) and $\sim$4 Gyr.

\subsubsection{Other Age Indicators \label{sec:otherage}}

We examine other age diagnostics for HD 984 to help constrain its age. HD 
984 appears to be fairly chromospherically and coronally active, which hint 
at fast rotation and youth. \citet{Wright04} estimated an age for HD 984 of 
0.49 Gyr based on chromospheric activity. The star has mass 
$\sim$1.2 M$_{\odot}$, B-V $\simeq$ 0.52, and 
T$_{\rm eff}$ $\simeq$ 6300\,K, which corresponds almost exactly with the 
``Kraft break'' \citep{Angus15,Kraft67}. This region divides F-type stars with 
thin convective envelopes, mostly radiative interiors with small convective 
cores, and which slow down less efficiently during their main sequence 
phases, from solar-type stars with thicker convective envelopes, radiative 
cores, and which slow down appreciably during their main sequence phase. 
The gyrochronology relations are not well-constrained near the Kraft break 
\citep[e.g.][]{Mamajek08} - and may not be applicable. We can, however, 
quantify just how active HD 984 is compared to nearby stars of similar 
spectral type.

{\it Coronal X-ray and Chromospheric Ca H \& K Emission:} HD 984 has an 
X-ray counterpart (1RXS J001410.1-071200) in the ROSAT All-Sky Survey 
Bright Source Catalog \citep{Voges99}. The X-ray source is 4'' away from 
the optical position, but with position error of 8'', the match with HD\,984 is 
highly probable. Based on the ROSAT X-ray count rate 
(0.265\,$\pm$\,0.029 ct\,s$^{-1}$) and hardness ratio HR1 
(HR1 = -0.20\,$\pm$\,0.10), and using the formulae in \citet{Fleming95} and 
\citet{Mamajek08}, we estimate the following properties: 0.1-2.4 keV X-ray 
flux $f_X$ = 1.92\,$\times$\,10$^{-12}$ erg\,s$^{-1}$\,cm$^{-2}$, log(L$_X$/erg/s)
= 29.71 dex, and log(L$_X$/L$_{bol}$) = -4.22 dex. It is unlikely that the 
X-ray emission is from the low-mass companion alone as the star would 
have to be emitting $\sim$40$\times$ more powerful than that predicted 
for saturated X-ray emission (log(L$_X$/L$_{bol}$), so the primary is 
almost certainly completely dominating the X-ray flux. Among nearby F-type 
dwarfs in the ROSAT study of \citet{Huensch99}, the mean relation of 
coronal X-ray activity values goes as log(L$_{X}$/L$_{bol}$) $\simeq$ 
-6.60 + 2.67(B-V), with rms scatter $\pm$0.68 dex. HD\,984 is 0.98 dex 
($\sim$1.4$\sigma$) brighter in log(L$_X$/L$_{bol}$) than this locus and 
suggestive that the star is amongst the most active $\sim$4\%\, of F-type 
dwarfs.

Based just on the coronal X-ray emission (log(L$_X$/L$_{bol}$ = -4.22 dex), 
one would predict strong chromospheric activity 
log(R$^{\prime}_{HK}$) $\simeq$ -4.34 using equation A1 of \citet{Mamajek08}.
Indeed, this predicted log(R$^{\prime}_{HK}$) is close to that reported from 
multiple surveys -- log(R$^{\prime}_{HK}$) ranges from -4.31 
\citep{Schroeder09} to -4.46 \citep{Pace13}, with a mean value of 
approximately -4.40 \citep{Wright04,White07,Schroeder09,Isaacson10,
Pace13}. In a catalog of unique FGK stars with log(R$^{\prime}_{HK}$) 
values from the surveys of \citet{Henry96}, \citet{Wright04}, and \citet{Gray06}, 
one finds 146 single main sequence Hipparcos stars with B-V within 
$\pm$0.02 mag of HD 984's color (0.52). Among these dwarf stars of similar 
color to HD 984, its log(R$^{\prime}_{HK}$) value ($\sim$ -4.40) places it 
among the most chromospherically active 5\%, similar to that inferred from 
the coronal activity. The star's combination of color (B-V = 0.52) and 
chromospheric activity (log(R$^{\prime}_{HK}$) $\simeq$ -4.40) for HD\,984 
can also be compared to stars to various samples in Fig. 4 of \citet{Mamajek08}: 
Sco-Cen ($\sim$0.01-0.02 Gyr), the Pleiades ($\sim$0.13 Gyr), the Hyades 
($\sim$0.6 Gyr), the M67 cluster ($\sim$4 Gyr), and field stars. The star's 
chromospheric activity is less active than typical Sco-Cen members, most 
consistent with the Pleiades locus, would be near the upper limit of that seen 
among Hyades members, and is way more active than M67 members and the 
median values calculated for late F-type stars.

{\it Rotation:} Given the star's projected rotational velocity 
\citep[$v$sin$i$ = 42.13\,$\pm$\,1.65 km\,s$^{-1}$;][]{White07} and the 
calculated radius (1.247\,$\pm$\,0.053 R$_{\odot}$), this implies that the star's 
rotation period must be P$_{rot}$ $<$ 1.5 day. Note that among 487 stars 
classified as F7 dwarfs in the \citet{Glebocki05} compendium of projected 
rotational velocities ($v$sin$i$), HD\,984 is among the top $\sim$1\%\, fastest 
rotating stars. The F7 dwarfs in the \citet{Glebocki05} catalog have median 
$v$sin$i$ $\simeq$ 7.2 km\,s$^{-1}$ with 95\%\, range of 2.7-35.2 km\,s$^{-1}$.
The mean $v$sin$i$ value for a F7 star of similar color and absolute magnitude 
to HD\,984 in the $\sim$0.6 Gyr-old Hyades cluster is approximately 
12\,km\,s$^{-1}$, and none within $\pm$0.05 mag in B-V of HD\,984 have 
$v$sin$i$ $>$ 20 km\,s$^{-1}$ \citep{Soderblom93}. In the Pleiades, however, 
there are stars of similar color to HD\,984 with $v$sin$i$ of $\sim$40-60 
km\,s$^{-1}$ \citep{Soderblom93}. The star's combination of 
log(L$_X$/L$_{bol}$) and $v$sin$i$ for its V-I color \citep[0.59;][]{ESA97} 
appears typical among F-type stars in the $\sim$50 Myr-old IC 2391 and IC 
2602 clusters, and would place it among the fastest rotators in the 
$\sim$130 Myr-old Pleiades \citep[see Fig. 6 of ][]{Stauffer97}, where the 
quoted ages are Li-depletion ages from \citet{Barrado04} and \citet{Dobbie10}.

The rotation, coronal activity, and chromospheric activity indicators for HD 984 
seem consistent with a very young, active main sequence star. The rotation 
and X-ray emission are consistent with the star being amongst the youngest 
$\sim$1-5\%\, of F7 dwarfs. The main sequence lifetimes of 
$\sim$1.2 M$_{\odot}$ stars are $\sim$5 Gyr \citep{Bressan12}, suggesting 
that if HD 984 is amongst the youngest $\sim$1-4\%\, of F7 dwarfs, that its age 
is very likely to be $\lesssim$200 Myr. This would be consistent with its 
chromospheric activity being similar to that of Pleiades members, its rotation 
being similar to that of $\sim$50 Myr stars in IC 2391 and IC 2602 (although 
near the upper limit of projected rotation velocities for Pleiades members), and
its rotation and activity indicators all being very inconsistent with Hyades-age 
stars. Based on rotation and activity indicators, we adopt a strong upper limit 
on the age of HD 984 of $<$200 Myr. The fact that the star is on the main 
sequence places an approximate lower limit of 30 Myr. We conclude that an 
approximate age for the primary star HD\,984 that is independent of any 
kinematic membership or consideration of the properties of the secondary is 
30-200 Myr (95\%CL).

\subsection{Companion Characteristics}
\label{sec:Mass}

\begin{figure}
\centering
\includegraphics[width=84mm]{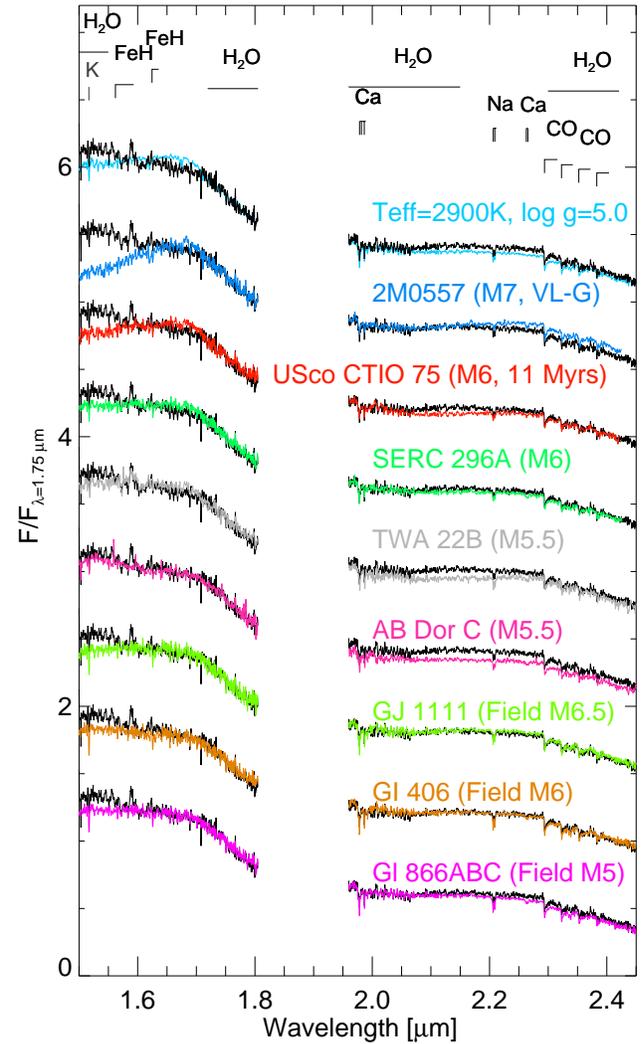}
\caption{Comparison of HD 984 B spectrum (black) to those of field dwarfs, 
young companions, isolated objects, and to the best-fitting BT-COND 
spectrum. It enables us to conclude that HD 984 B is an M$6.0\pm0.5$ dwarf.}
\label{spectrum}
\end{figure}

\begin{figure}
  \centering
    \includegraphics[width=84mm]{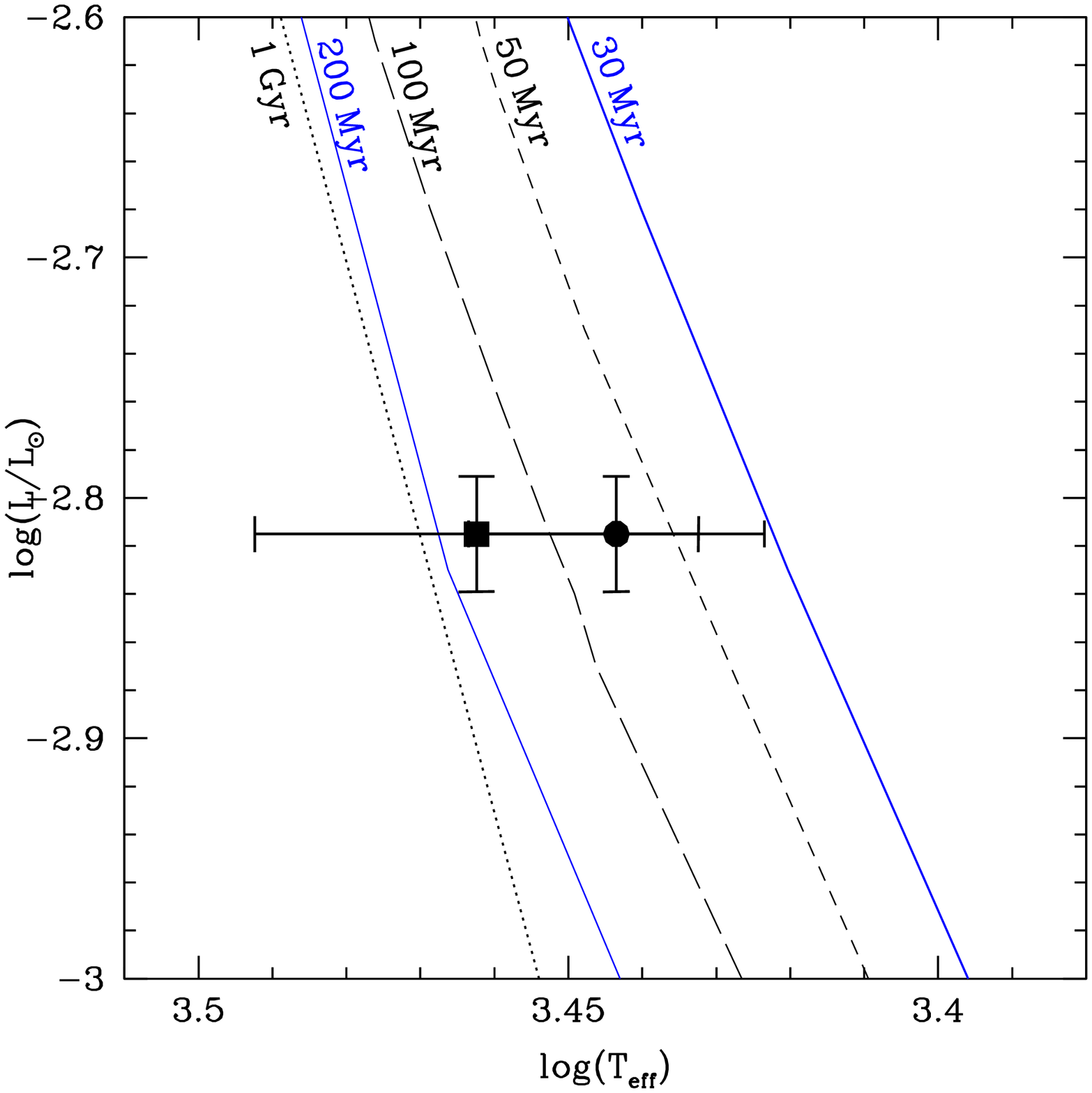}
  \caption{HR diagram position of HD 984 B in bolometric luminosity (right) 
  based on our spectral type estimate and adopting the \citet{Stephens09} 
  T$_{\rm eff}$ scale (circle), BT-COND and GAIA synthetic spectra 
  (\citet{Brott05, Allard13}: square). The BHAC15 \citep{Baraffe15} evolutionary 
  tracks are plotted for 30, 50, 100, 200 Myr and 1 Gyr. The 30 and 200 Myr 
  isochrones are in blue to indicate the consistency with the lower and upper 
  age limits determined in this work. }
  \label{hr_companion}
\end{figure}

\begin{table}
\caption{System properties} 
\begin{center} 
\begin{tabular}{p{0.25\linewidth} p{0.21\linewidth} p{0.15\linewidth} p{0.2\linewidth}}
\hline 
   Property & HD 984 &  & HD 984 B  \\
  \hline
   Distance (pc)$^{\rm a}$ & & 47.1\,$\pm$\,1.4 & \\ 
   Age (Myr)$^{\rm b}$  & & 30-200& \\ 
   $A_{V}$ $^{\rm c}$ & & 0& \\    
   $T_{\rm eff}$  & 6315\,$\pm$\,89 & & $2777^{+127}_{-130}$ $^{\rm d}$\\   
     & & & $2900\pm200$ $^{\rm e}$  \\
   Spectral type  & F7V && M$6.0\pm0.5$ \\   
   log($L$/$L_{\odot}$) & 0.36\,$\pm$\,0.03 & & -2.815$\pm$0.024 \\ 
   Mass & 1.20\,$\pm$\,0.06 M$_{\odot}$  & & $74^{+27}_{-31}$ $M_{\rm Jup}$ $^{\rm d}$\\
   & & & $90^{+69}_{-42}$ $M_{\rm Jup}$ $^{\rm e}$\\
   & & & $84^{+29}_{-42}$ $M_{\rm Jup}$ $^{\rm f}$\\
   Separation (mas)  &&  $201.6\pm0.4$ & \\ 
   P.A.($^{\circ}$) & & $92.2\pm0.5$  & \\ 
   $H$  & 6.170\,$\pm$\,0.023&& 12.58\,$\pm$\,0.05\\   
   $Ks$  & 6.073\,$\pm$\,0.038 && 12.19\,$\pm$\,0.04 \\ 
   $L'$ & 6.0\,$\pm$\,0.1 & & 12.0\,$\pm$\,0.2 \\ 
\hline
   \end{tabular} 
  \end{center}
{\bf $^a$} {\textit{Hipparcos} catalog \citep{vanLeeuwen07}.}\\
{\bf $^b$} {This work.}\\
{\bf $^c$} {\citet{Schlegel98} and \citet{Reis11}.}\\
{\bf $^d$} {Based on the spectral type conversion scale of \citet{Stephens09}.}\\
{\bf $^e$} {Based on BT-COND and GAIA synthetic spectra.}\\
{\bf $^f$} {Converted from the derived log($L$/$L_{\odot}$) for HD 984 B.}\\
\label{table:properties}
\end{table}

Based on the NaCo $L'$ photometry alone, the companion mass can only 
be estimated from theoretical models which are highly dependent on the 
age of the star and lead to large uncertainties in the determined mass. 
Using the COND evolutionary tracks \citep{Baraffe03}, if the system is 30 
Myr, the companion may be a $33\,\pm\,6 M_{\rm Jup}$ young brown dwarf. 
If the system is $200$ Myr, the companion may be a $0.09\pm0.01 
M_{\odot}$ low mass star, likely an M5-M6 dwarf. The errors are based on 
the uncertainty in the photometry and the distance of the star and do not 
include systematic uncertainties in the models. The range of companion 
masses based on the DUSTY evolutionary model \citep{Chabrier00} is 
within the errors of the values derived from the COND model: 30 Myr and 
$34\,\pm\,6 M_{\rm Jup}$ or 200 Myr and $0.10\pm0.01 M_{\odot}$. Thus, 
the companion mass estimation based on photometry is not significantly 
impacted by the evolutionary model chosen.

We compare our CADI analysis of the SINFONI spectrum (black spectrum, 
see \autoref{spectrum}) with field dwarfs from the IRTF library 
\citep{Rayner09}. We perform a least squares fit of the spectra to determine 
the best fit. In addition to the overall spectral slope from 1.5 to 2.45 $\mu$m, 
we aim to fit the following spectral features: the 
\ensuremath{\mathrm{K}}\,\textsc{i} band at 1.516 $\mu$m, the 
\ensuremath{\mathrm{Ca}}\,\textsc{i} triplet near 1.98 $\mu$m, the 
\ensuremath{\mathrm{Na}}\,\textsc{i} doublet near 2.207 $\mu$m, the 
\ensuremath{\mathrm{CO}} overtones longward of 2.3 $\mu$m, and the 
overall shape of the $Ks$-band which is sensitive to the collision-induced 
absorption of molecular hydrogen, thus to the atmospheric pressure and 
surface gravity. The companion has features later than M5 and midway 
between those of M5.5 and M6.5 field dwarfs. The SINFONI spectra of the 
young ($\leq$150 Myr) mid-M companions AB Dor C and TWA 22 B 
\citep{Close07, Bonnefoy09} reproduce the pseudo-continuum in the 
$H$-band, but have bluer slopes than our companion, suggesting a later 
spectral type for HD 984 B. Spectra of other young mid-M dwarfs have a 
more triangular $H$-band shape than our object. The companion fits well 
with the M6 object SERC 296A \citep{Thackrah97,Allers13}, which does not 
appear to be a good candidate member to any nearby associations based on 
its kinematics \citep{Gagne14}, but has an age below $\sim$ 200 Myr due to 
lithium absorption. We therefore conclude that the HD 984 B is more likely a 
M$6.0\pm0.5$ object, which is younger than the typical field dwarf age 
($>>$1 Gyr). The subtype accuracy is due to the companion features being 
intermediate between a M5.5 and an M6.5.

The spectral type of M$6.0\pm0.5$ corresponds to 
T$_{\rm eff}=2777^{+127}_{-130}$ K using the conversion scale of 
\citet{Stephens09}. The T$_{\rm eff}$ accuracy is the quadratic combination of 
the spectral type estimate and systematic uncertainty. Conversely, the 
companion's pseudo-continuum shape and main absorption features are best 
reproduced by BT-COND and GAIA synthetic spectra \citep{Brott05, Allard13} 
with T$_{\rm eff}=2900\pm200$ K and log g=5.0-5.5 dex. The accuracy on 
T$_{\rm eff}$ is limited by the intrinsic differences and uncertainties between 
the atmospheric model grids ($H$-band is known to be badly reproduced by 
models, see \citealt{Bonnefoy13}). \citet{Rajpurohit13} find a T$_{\rm eff}$ 
range of 2700 to 3000 K for M5.5 to M6.5 dwarfs by fitting BT-Settl 
\citep{Allard13} spectra to optical spectra of M-dwarfs. This is consistent with 
our spectral type and derived temperature. 

We calculate the bolometric luminosity based on our derived spectral type 
and $H$, $Ks$ and $L'$-band companion magnitude measurements. We adopt 
bolometric corrections of $BC_{H(B)}= 2.588\pm0.032$ mag, 
$BC_{Ks(B)}= 2.940\pm-0.015$ mag, and $BC_{L'(B)}= 3.260\pm0.022$ mag 
by interpolating between an M6 and M7 dwarf, using bolometric correction and 
color information for M dwarfs from \citet{Pecaut13}, \citet{Schmidt14},  and 
\citet{Dupuy12}. By taking the weighted mean of our bolometric luminosity 
calculations from $H$, $Ks$ and $L'$-band measurements, we find 
log(L/L$_{\odot}$)=-2.815$\pm$0.024 dex. These results are summarized in 
Table \ref{table:properties}.

\autoref{hr_companion} compares the derived T$_{\rm eff}$ of the companion 
against the BHAC15 \citep{Baraffe15} evolutionary tracks of different ages. 
Comparing our derived bolometric luminosity with evolutionary tracks suggests 
that the companion is consistent with our derived stellar age of 115\,$\pm$\,85 
Myr. HD 984 B demonstrates the challenges in fitting M-dwarfs on low mass 
evolutionary tracks. Thus, we can rule out ages less than 30 Myr based 
on the companion HR diagram position and we have demonstrated that this 
companion is consistent with an M$6.0\pm0.5$ object.

Recently, \citet{Hinkley15} announced the detection of seven companions with 
short projected separations around intermediate mass stars in the 
Scorpius-Centaurus association. HD 984 B has a companion-to-star mass ratio 
of 0.067$^{+0.028}_{-0.035}$ based on the mass derived from the luminosity. 
Given the small separation of the system (9.5 AU), HD 984 B shares similar 
characteristics as the aforementioned companions discovered by \citet{Hinkley15}.

\section{Conclusion}
\label{sec:Conclusion}
We report the discovery of a low-mass companion to the F7V star HD 984. This 
companion was detected in $L'$-band with the Apodizing Phase Plate 
coronagraph and non-coronagraphic photometry with the NaCo instrument on the 
VLT. HD 984 has been reported to be part of the 30 Myr old Columba association. 
However, based on our independent analysis of several age indicators, we 
estimate a main sequence age of 115\,$\pm$\,85 Myr (95\% CL). Due to the age 
uncertainty, the companion mass may range from a low mass brown dwarf 
($\sim$33 M$_{\rm Jup}$) to an M dwarf ($\sim$0.09 M$_{\odot}$), using the 
COND evolutionary models.

We analyze the slope and shape of SINFONI $H+K$ IFU data of the companion 
compared with field M dwarfs. We conclude the companion is an 
M$6.0\pm0.5$ dwarf. Using our derived spectral type, we aim to determine the 
age of the system by placing the companion $L'$-band absolute magnitude and 
bolometric luminosity on COND evolutionary tracks. While the $L'$-band HR 
diagram position allows us to rule out an age less than 50 Myr, the companion's 
bolometric luminosity position is consistent with an age of $\geq$30 Myr. Thus, 
we cannot set age constraints on the companion, due to this discrepancy 
between the companion's position on evolutionary tracks. Future observations in 
$J$-band could help to look for signatures of low surface gravity in HD 984 B.

Given its small projected separation of just $\sim$9 AU, HD 984 B will show 
significant orbital motion over the next few years, allowing the potential for 
dynamical mass determination. In order to determine the individual mass 
components of the HD 984 system, RV data must be obtained. Using the current 
projected separation as an approximation for the semi-major axis and assuming 
a circular orbit, we would expect to detect a maximum semi-amplitude from RV 
of 1.01 km/s for an M-dwarf companion and 0.27 km/s for a brown dwarf. This 
calculation assumes the orbit is edge-on ($i$=90$^{\circ}$), however based on 
our two epoch astrometric  measurements, its orbit is unlikely to be edge-on. 
Even if the companion is closer to face-on ($i$=5$^{\circ}$), the 
semi-amplitude of 0.087 km/s for an M-dwarf and 0.024 km/s for a brown dwarf is 
within the detection limits of RV. This system is also a promising target for GAIA, 
which could reveal the reflex motion of the star and assess the dynamical mass of 
the companion. These measurements are amplified if the system is face-on, 
complementing the RV method. Thus, the dynamical mass of this companion can 
be achieved with RV and astrometric measurements, providing a crucial 
comparison with the theoretical evolutionary models for mass determination.

We have demonstrated that, given the difficulty in deriving a reliable age for 
HD 984 that is independent of its purported group membership, the derived color 
and spectral parameters of the companion are necessary to determine the 
companion mass. These results suggest that caution should be used when 
estimating the masses of companions based on photometric data and stellar age 
based on kinematic group membership alone. It reinforces the importance of 
future near-infrared high contrast integral field spectrographs to the 
characterization of low-mass stellar and substellar companions.

\section*{Acknowledgments}
TM and MAK acknowledge funding under the Marie Curie International 
Reintegration Grant 277116 submitted under the Call FP7-PEOPLE-2010-RG. 
EEM acknowledges support from NSF award AST-1313029. Part of this work has 
been carried out within the frame of the National Centre for Competence in 
Research PlanetS supported by the Swiss National Science Foundation. SPQ 
and MRM acknowledge the financial support of the SNSF AML, GC, and JR 
acknowledge financial support from the French National Research Agency (ANR) 
through project grant ANR10-BLANC0504-01. This paper makes use of the 
SIMBAD Database and the Vizier Online Data Catalog.

\appendix
\section{SINFONI Orientation}
\label{Section:AppA}
We calibrated the absolute orientation of the field of view of SINFONI using 
observations of GQ Lup B from UT 2013 August 24 (technical program 
ID 60.A-9800). These observations were obtained using the pre-optics offering 
a 50$\times$100 mas sampling in the $H+K$ band. During the 54 min pupil 
tracking (PT) sequence, 100$\times$7s integrations were recorded. The field 
rotated by 12.84$^{\circ}$. We reduced these data following the same 
procedure as  HD 984 B. Once the cubes were corrected for atmospheric 
refraction, we removed the halo from the primary star centered in the field of 
view with a radial profile. 

We found that a clockwise rotation by the $ADA.POSANG + C$ values of each 
data cubes re-aligned the final GQ Lup data cubes with the North. 
$ADA.POSANG$ is a variable found in the image header, corresponding to the 
position angle of the rotator at the Cassegrain focus at the time of the 
observations. $C$ is an additional offset related to the calibration of the 
instrument rotator true North position.

We define $ROT.PT.OFF +C = 180^{\circ} + ADA.PUPILPOS + C$, the angular 
offset needed to realign the frames to the North when the parallactic angle is 0. 
$ADA.PUPILPOS$ is a keyword stored into the file header, which depends on 
the telescope pointing position and time of observation. It is redefined at the 
beginning of any PT sequence, but it remains constant during a PT observing 
sequence. 

We verified that this relation remains valid for other datasets obtained in PT 
mode on GQ Lup during the same night, on AB Dor C (UT 2013 October 17, 
60.A-9800), and on HD 984 B (UT 2014 October 10, 2014 December 3,5,8 
094.C-0719), and of the astrometric binary HD 179058 AB (UT 2014 April 26, 
60.A-9800). We note that we could not use the observations of the HD179058 
AB to properly calibrate the instrument plate scale and absolute orientation 
since the binary had likely moved on its orbit since its latest independent 
astrometric measurement \citep{Tokovinin10}. 

For the case of GQ Lup B, the $ADA.PUPILPOS$ keyword was fixed to 
-0.07297. Therefore, we adopted ROT.PT.OFF=179.927. We estimated a 
value of C=$0.0 \pm 0.5^{\circ}$ comparing the resulting position angle in the 
derotated cubes to the position angle of the system measured from VLT/NaCo 
data obtained on 2012 March 3 and reported in \citet{Ginski14}. We note that 
this value of C assumes that the companion did not have significant orbital 
motion in the course of one year. This is reasonable given the available 
VLT/NaCo astrometry of the system recorded since 2008 
\citep[see Table 2 of][]{Ginski14}. We also deduce from the SINFONI data of 
GQ Lup B that the mean square plate scale is $49.30 \pm 0.14$ mas/spaxel 
when the 50$\times$100 mas and $H+K$ band mode of the instrument are 
chosen.

\bibliographystyle{apj}  

\label{lastpage}

\end{document}